\documentclass[runningheads]{llncs}
\usepackage{graphicx}
\usepackage{multirow}
\usepackage{todonotes}
\usepackage{hyperref}

\makeatletter
\newcommand{\printfnsymbol}[1]{%
  \textsuperscript{\@fnsymbol{#1}}%
}
\makeatother

\begin{document}
\title{Late fusion of deep learning and hand-crafted features for Achilles tendon healing monitoring}
\author{Norbert Kapinski\inst{1}\printfnsymbol{1}, Jedrzej M. Nowosielski\inst{1}\thanks{equal contribution}, Maciej E. Marchwiany\inst{1}, Jakub Zielinski\inst{1}, Beata Ciszkowska-Lyson\inst{3}, Bartosz A. Borucki\inst{1}, Tomasz Trzcinski\inst{2,4}, Krzysztof S. Nowinski\inst{1}}

\institute{$^1$University of Warsaw $^2$Warsaw University of Technology\\
$^3$Carolina Medical Center $^4$Tooploox}

\maketitle              
\begin{abstract}
Healing process assessment of the Achilles tendon is usually a complex procedure that relies on a combination of biomechanical and medical imaging tests. As a result, diagnostics remains a tedious and long-lasting task. Recently, a novel method for the automatic assessment of tendon healing based on Magnetic Resonance Imaging and deep learning was introduced. The method assesses six parameters related to the treatment progress utilising a modified pre-trained network, PCA-reduced space and linear regression. In this paper, we propose to improve this approach by incorporating hand-crafted features. We first perform a feature selection in order to obtain optimal sets of mixed hand-crafted and deep learning predictors. With the use of approx. 20,000 MRI slices, we then train a meta-regression algorithm that performs the tendon healing assessment. Finally, we evaluate the method against scores given by an experienced radiologist. In comparison with the previous baseline method, our approach significantly improves correlation in all of the six parameters assessed. Furthermore, our method uses only one MRI protocol and saves up to 60\% of the time needed for data acquisition.

\end{abstract}
\section{Introduction}
Achilles tendon rupture is common among physically active middle-aged people. It seriously affects the patient's mobility and ability to be physically active over a long period of time. Proper oversight during the rehabilitation is important and can lead to a reduction of complications i.a. tendon re-rupture.

Existing methods like ATRS \cite{Kearney2012} are only suitable for measuring the general outcome of the rehabilitation, related to symptoms and physical activity of patients. The inclusion of medical imaging allows to complement the monitoring and properly assess tissue morphology associated with the tendon state. However, due to costs of medical examinations, limited time and resources of radiology departments, it is still an uncommon approach.

Both problems were recently addressed in~\cite{Kapinski2018}, where the authors presented a first MRI-based method for automatic assessment of the Achilles tendon healing progression as well as selected the two most informative protocols suitable for their approach. The method is treated as a state-of-the-art baseline for our studies.

We aim to improve the method in terms of the assessment quality as well as the number of MRI protocols required, hence enabling clinics to perform more studies and efficiently assist radiologists in patients evaluation. To do so, we incorporate hand-crafted features to the previous approach and perform a late fusion with the use of a meta-regression algorithm. More precisely, the baseline method consists of a convolutional neural network feature extractor and a principal component layer that performs dimensionality reduction. We take 200 first principal components and 46 hand-crafted features investigated in~\cite{Nowosielski17}. Subsequently, we select an optimal mixed feature set with the use of the LASSO method. Finally, we train a meta-regression algorithm to fit the resulting representation to the 6 tendon state scores assigned by a human annotator. Our method outperforms the baseline model in terms of correlation with the ground truth in every single parameter as well as improve the mean absolute error for 4 out of 6 and max absolute error for 2 out of 6 parameters. Furthermore, the proposed method uses only one MRI protocol instead of two, which directly translates to up to a 60\% shorter time of acquisition and lower cost.  

\section{Method}
In this section, we describe our improved method for predicting the Achilles tendon healing phase. We start by selecting the most valuable representation of features for training our meta-regression algorithm. To do so we apply the LASSO method on a set of 246 features, being a combination of the 200 most significant principal component features extracted by a baseline model and 46 hand-crafted predictors. The latter represents statistics from the Region of Interest (ROI) which in our case is the segmented tendon. More precisely, the new features include the ROI area, pixel value based statistics over the ROI (min, max, mean, standard deviation, skewness, kurtosis, 25-percentile, median, 75-percentile), as well as Haralick's textural features~\cite{Haralick1973,Nowosielski17} (angular second moment, contrast, correlation, variance, inverse difference moment, sum average, sum variance, sum entropy, entropy, difference variance, difference entropy, maximum probability) for separation distance \textit{d}=1,5,10. 

We use the resulting representation of mixed features after LASSO transformation and further described ground truth labels to train a regression algorithm. Finally, we use the $H$ metric proposed in the baseline method \cite{Kapinski2018} to represent a score of the Achilles tendon condition, visible in a single 3D MRI study:
\begin{equation}
H = TM(R(x_1), R(x_2),..., R(x_n))
\end{equation}
where $TM$ is a truncated mean with 2.5 upper and lower hinges (a value used by the baseline method), $R(x_i)$ is the regression score computed on the slice $x_i$ where $i$ is the index of the slice in the 3D MRI study. 

\begin{figure}
\includegraphics[width=\textwidth]{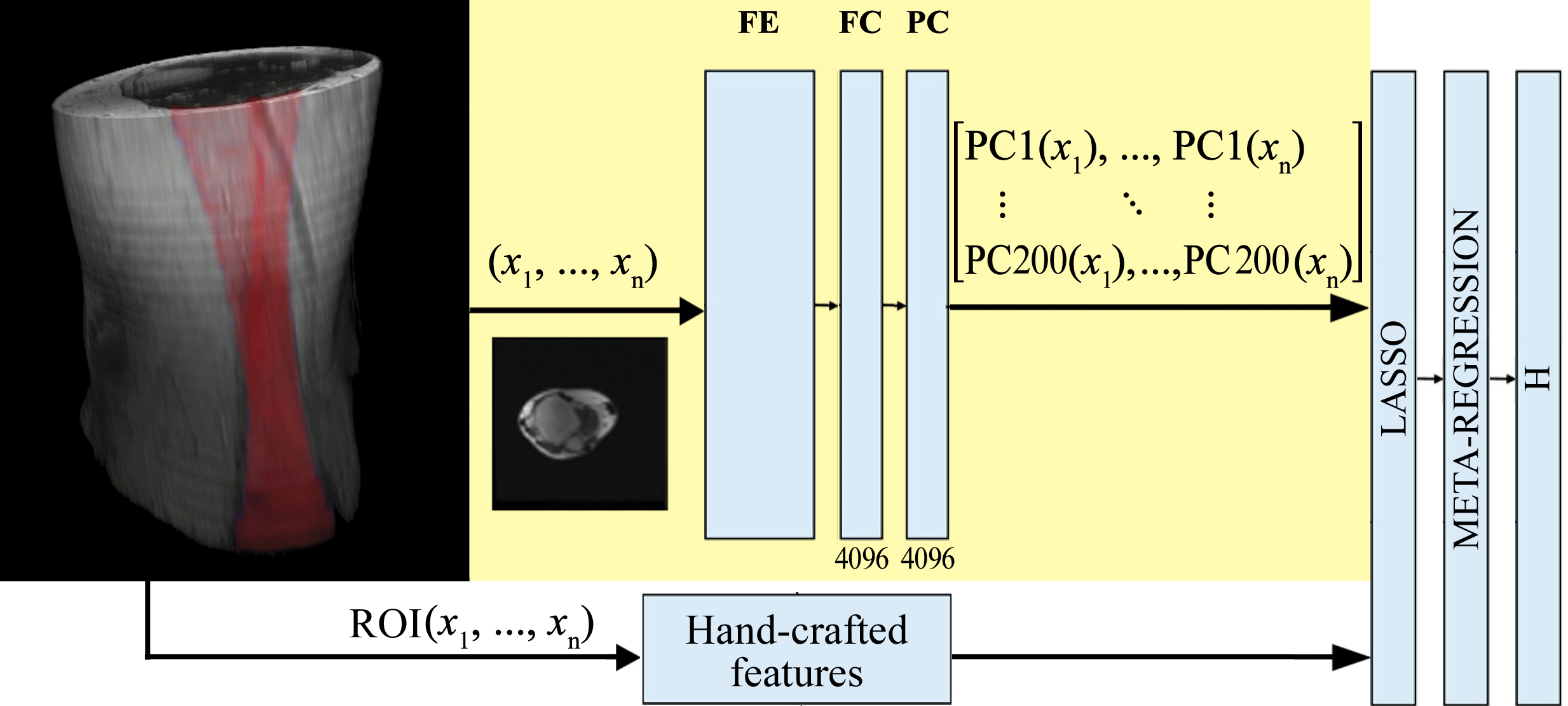}
\caption{The late fusion algorithm topology used to generate tendon healing assessment.} \label{fig:net}
\end{figure}

Fig.~\ref{fig:net} shows the overview of our framework based on a late fusion of so-called deep learning features and hand-crafted ones.
In general, the approach for each 2D axial slice in the 3D MRI study produces a single assessment score that is merged into one value with the use of the $H$ metric. 

The deep learning features (without ROI information) are computed in the same way as described in the baseline paper (see region marked in yellow on Fig.~\ref{fig:net}). It is a truncated version of the AlexNet convolutional neural network~\cite{AlexNet} trained with the use of 10 MRI protocols on a binary task of distinguishing between healthy and injured tendons. The architecture consists of the feature extractor, the first fully connected layer (namely fc6) followed by a principal component layer that performs dimensionality reduction. According to the authors of this approach, the first 200 principal components preserve a 98.8\% amount of variance from 4096 fc6 activation outputs. Thus, we use all 200 as inputs to our late fusion approach. The model and the principal component transformation remain the same as in the original paper. 

On top of the previous approach, we developed the novel contribution of this paper, namely the meta-regression model that combines information conveyed by the hand-crafted and deep learning features. The approach uses the LASSO feature selection, that allows us to reduce dimensionality and then effectively train the meta-regression without overfitting. 

\section{Experiments}
Within this section, we present experiments that allowed us to select the components for our final method. We start by introducing our dataset. Next, we show an analysis of different MRI protocols that brought us to the selection of the final input data. Subsequently, we present the feature selection method followed by a detailed study on training different meta-regression algorithms. We conclude this section by comparing the results obtained by our method and the baseline approaches for 4 test patients excluded from the training procedure.

\subsection{Dataset}
The acquired dataset includes 3D MRI scans of 60 injured patients that suffered from an acute Achilles tendon rupture. The injured patients have their lower limb scanned once before the surgical reconstruction and then 9 times after the surgery, i.e., 1, 3, 6, 9, 12, 20, 26, 39, 52 weeks after the operation (10 MRI studies altogether for a single injured patient). Healthy volunteers were scanned only once. The single MRI study includes scans performed with 10 MRI protocols i.e four 3D FSPGR Ideal [Fast Spoiled Gradient Echo] (In Phase, Out Phase, Fat, Water), PD [Proton Density], T1, T2, T2 mapping, T2$^\ast$ GRE [Gradient Echo] and T2$^\ast$ GRE TE\_MIN [Minimal Time Echo].

Within our dataset, there is the complete sequence of 10 MRI studies in time (including the manual segmentation and the ground truth for the radiological survey) available for 48 injured patients. In the case of T2$^\ast$ GRE TE\_MIN (our selected protocol based on further described studies), these 48 patients translate to 480 MRI studies and 18,863 slices with non-empty tendon ROI. We randomly selected 4 patients (40 studies and 1545 slices) to form a separate test set. The remaining 44 injured patients constitute a training set which is used for cross-validation and for the final training of the feature selector and the meta-regression. 

Manual segmentation of the Achilles tendon ROI has been provided by an expert radiologist, who also annotated the ground truth parameters through the scoring of several aspects of the tissue, namely: \begin{enumerate}
\item Structural changes within the tendon (SCT)
\item Tendon thickening (TT)
\item Sharpness of the tendon edges (STE)
\item Tendon edema (TE)
\item Tendon uniformity (TU)
\item Tissue edema (TisE)
\end{enumerate}
The TisE parameter relies on the tissues outside the tendon and the STE parameter on the border, thus both can be treated as extra-tendon scores. All of the other parameters rely on intra-tendon structures. The assessment of the parameters was done on a scale ranging from $1$ to $7$, where $1$ represents a healthy tendon and $7$ a severely injured one.  

\subsection{Tendon healing assessment with late fusion}
\subsubsection{MRI protocol selection:} 
\begin{figure*}[h]
\centering
\includegraphics[width=1.0\textwidth]{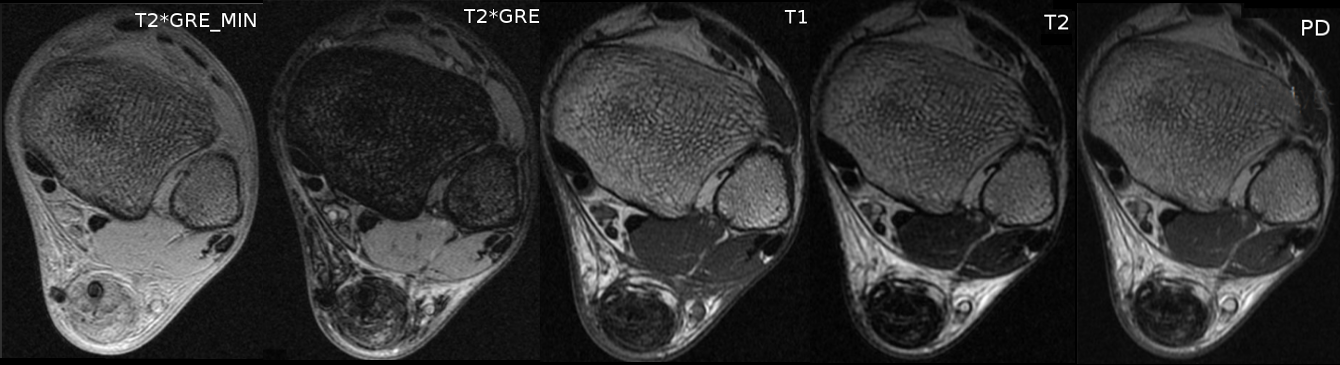}
\caption{Comparison of different MRI protocols illustrating the healing process of the Achilles tendon in the 12 week following reconstruction.} \label{fig:protocol-selection}
\end{figure*}

Within this study we focus on selecting one MRI protocol that shows the most valuable information regarding the tendon healing process. We discovered that the MRI signal from the tendon area in the subsequent healing weeks is more differentiable with the use of the T2$^\ast$GRE TE\_MIN protocol, than with the use of any other (see Fig \ref{fig:protocol-selection}). 

As healthy tendon tissue is characterised by very short T2 times, the tendon visible on the images derived from the other protocols is almost black after just 12 weeks of rehabilitation, while the T2$^\ast$GRE TE\_MIN still detects pathological changes. Better sensitivity of this protocol for the partially healed Achilles tendon results from its very short echo time. Considering this observation, we decided to use T2$^\ast$GRE TE\_MIN as the input data for our method.

\subsubsection{Meta-regression training task:}
We followed a standard 4-fold cross-validation procedure to tune the $\alpha$ hyper-parameter of the LASSO feature selector and hyper-parameters of meta-regression algorithms. We have chosen 4 groups to ensure approximately equally sized folds. Each fold contains slices from 11 patients (approx. 4300 axial slices).

In terms of the LASSO method we obtained best results regarding correlation with the ground truth for $\alpha=0.1$ and the following feature set: 
\begin{itemize}
\item SCT:  6 DL features and 10 hand-crafted features, including 3 Haralick's features (sum variance for d=1, sum average for d=5, sum average for d=10);

\item TT: 5 DL features and 7 hand-crafted features, but no Haralick's features;

\item STE: 5 DL features and 8 hand-crafted features, but no Haralick's features;

\item TE: 4 DL features and 8 hand-crafted feature, with one Haralick's feature (sum average for d=10);

\item TU: 4 DL features and 9 hand-crafted features, with one Haralick's feature (sum variance for d=1);  

\item TisE: 6 DL features and 7 hand-crafted features, but no Haralick's features.
\end{itemize}
The total number of selected features is always below 20 for all of the ground truth parameters. Furthermore, there are both deep learning and hand-crafted features present in all of the sets, indicating a possibility of successful fusion approach.

Using the limited feature sets, for each of the ground truth parameters we train several regression algorithms, namely: linear regression (LR), second degree polynomial regression (poly), support vector regression (SVR), multi-layer perceptron regression with 4 units in the hidden layer (MPR) and random forest (RF). Despite our multiple trials with different random forest sizes, the algorithm always showed a tendency to overfitting, hence we exclude RF from the table presenting meta-regression algorithms performance on the training set (Tab. \ref{tab:trainset}).

\begin{table*}[t]
\caption{Train set results for the tendon healing process monitoring.}
\begin{center}
\begin{tabular}{lc||c|c|c|c|c|c}
\textbf{Model} & & \textbf{SCT} & \textbf{TT} & \textbf{STE} & \textbf{TE} & \textbf{TU} & \textbf{TisE}\\ 

    \hline
    \multirow{3}{*}{poly}
    & MAE & $1.00\pm0.03$ & $0.62\pm0.02$ & $0.73\pm0.02$ & $0.76\pm0.02$ & $0.89\pm0.03$ & $0.70\pm0.02$\\
    & MAX-AE & 3.53 & 2.35 & 3.62 & 2.49 & 2.90 & 2.64 \\
    & Corr   & 0.87 & 0.82 & 0.46 & 0.80 & 0.65 & 0.87 \\
    \hline
    \multirow{3}{*}{SVR}
    & MAE & $0.88\pm0.01$ & $0.59\pm0.01$ & $0.67\pm0.01$ & $0.69\pm0.01$ & $0.83\pm0.01$ & $0.63\pm0.01$\\
    & MAX-AE & 3.73 & 2.32 & 3.83 & 2.50 & 2.95 & 2.75 \\
    & Corr   & 0.89 & 0.85 & 0.59 & 0.83 & 0.72 & 0.88 \\
    \hline
    \multirow{3}{*}{LR}
    & MAE & $1.00\pm0.04$ & $0.62\pm0.02$ & $0.74\pm0.02$ & $0.74\pm0.02$ & $0.90\pm0.03$ & $0.69\pm0.02$\\
    & MAX-AE & 3.52 & 2.39 & 3.77 & 2.52 & 2.88 & 2.65 \\
    & Corr   & 0.87 & 0.83 & 0.46 & 0.80 & 0.65 & 0.87 \\
    \hline
    \multirow{3}{*}{MPR}
    & MAE & $1.00\pm0.04$ & $0.63\pm0.02$ & $0.74\pm0.02$ & $0.71\pm0.02$ & $0.88\pm0.03$ & $0.96\pm0.02$\\
    & MAX-AE & 3.47 & 2.51 & 3.57 & 2.52 & 2.86 & 2.65 \\
    & Corr   & 0.86 & 0.83 & 0.46 & 0.80 & 0.65 & 0.88 \\
    \hline
    \end{tabular}
\end{center}
\label{tab:trainset}
\end{table*}

The models are evaluated with three metrics, i.e. mean absolute error (MAE), maximal absolute error (MAX-AE) and Fisher Z-Transformed mean Pearson correlations between computed values and ground truth scores for an individual patient. Presented meta-regression approaches resulted in comparable scores, thus we select them all in the following experiments. 

\subsubsection{Healing progress assessment:}

In this task, we evaluate our late fusion approach against the baseline. We use two variants of the baseline introduced in the original paper: (1) inference based on PD, T2$^\ast$ GRE and T2$^\ast$ GRE TE\_MIN protocols and (2) inference based only on the T2$^\ast$ GRE TE\_MIN protocol. In (1) we focus on reproducing the pipeline as in the original paper and (to make a fair comparison) enriched it with the protocol that we use, while in (2) we evaluate how the models inference on limited data only.

The test set performance for the meta-regression algorithms is summarised in Tab.~\ref{tab:testset}. 
\begin{table*}[t]
\caption{Test set results for the tendon healing process monitoring.}
\begin{center}
\begin{tabular}{lc||c|c|c|c|c|c}
\textbf{Model} & & \textbf{SCT} & \textbf{TT} & \textbf{STE} & \textbf{TE} & \textbf{TU} & \textbf{TisE}\\ 

    \hline
    \multirow{3}{*}{poly}
    & MAE & $1.15\pm0.13$ & $0.57\pm0.07$ & $0.75\pm0.08$ & $0.94\pm0.10$ & $0.92\pm0.09$ & $0.94\pm0.10$\\
    & MAX-AE & 2.67 & 1.78 & 1.81 & 2.50 & 2.12 & 2.39 \\
    & Corr   & 0.82 & 0.83 & 0.25 & 0.71 & 0.63 & 0.78 \\
    \hline
    \multirow{3}{*}{SVR}
    & MAE & $1.05\pm0.12$ & $0.56\pm0.06$ & $0.75\pm0.08$ & $0.91\pm0.10$ & $0.91\pm0.09$ & $0.94\pm0.10$\\
    & MAX-AE & 2.62 & 1.82 & 1.92 & 2.54 & 2.01 & 2.38 \\
    & Corr   & 0.85 & 0.85 & 0.31 & 0.72 & 0.65 & 0.80 \\
    \hline
    \multirow{3}{*}{LR}
    & MAE & $1.15\pm0.13$ & $0.55\pm0.06$ & $0.74\pm0.08$ & $0.90\pm0.10$ & $0.93\pm0.09$ & $0.97\pm0.10$\\
    & MAX-AE & 2.60 & 1.78 & 1.81 & 2.54 & 2.04 & 2.38 \\
    & Corr   & 0.84 & 0.84 & 0.18 & 0.71 & 0.62 & 0.78 \\
    \hline
    \multirow{3}{*}{MPR}
    & MAE & $1.13\pm0.12$ & $0.57\pm0.06$ & $0.74\pm0.08$ & $0.92\pm0.11$ & $0.91\pm0.09$ & $0.96\pm0.10$\\
    & MAX-AE & 2.63 & 1.77 & 1.78 & 2.54 & 2.04 & 2.40 \\
    & Corr   & 0.83 & 0.83 & 0.20 & 0.73 & 0.63 & 0.77 \\
    
    \hline
    \hline
    \multirow{3}{*}{baseline} 
    & MAE & $1.26\pm{0.07}$ & $0.71\pm{0.02}$ & $0.7\pm{0.03}$ & $0.97\pm{0.04}$ & $0.92\pm{0.05}$ & $0.99\pm{0.04}$ \\
    &MAX-AE & 3.53 & 2.49 & 1.91 & 2.34 & 2.2 & 2.47\\
    &Corr & 0.58 & 0.47 &-0.07 & 0.60 & 0.56 & 0.58\\
    \hline
    \multirow{3}{*}{\begin{tabular}{@{}c@{}} baseline \\ limited \end{tabular}} 
    & MAE & $1.24\pm{0.16}$ & $0.82\pm{0.09}$ & $0.75\pm{0.08}$ & $1.06\pm{0.10}$ & $0.90\pm{0.09}$ & $0.96\pm{0.10}$ \\
    &MAX-AE & 3.54 & 2.46 & 1.82 & 2.70 & 2.13 & 2.18\\
    &Corr   & 0.61 & 0.64 &-0.08 & 0.55 & 0.55 & 0.65\\
    \hline
    \end{tabular}
\end{center}
\label{tab:testset}
\end{table*}
In comparison with both baselines, even the simple linear meta-regression model allows to significantly improve correlation for all of the parameters, max absolute error for SCT, TT, STE and TU as well as mean absolute error for SCT, TE, and lastly TT as the one that has the statistical significance. 

\begin{figure*}[h]
\centering
\includegraphics[width=1.0\textwidth]{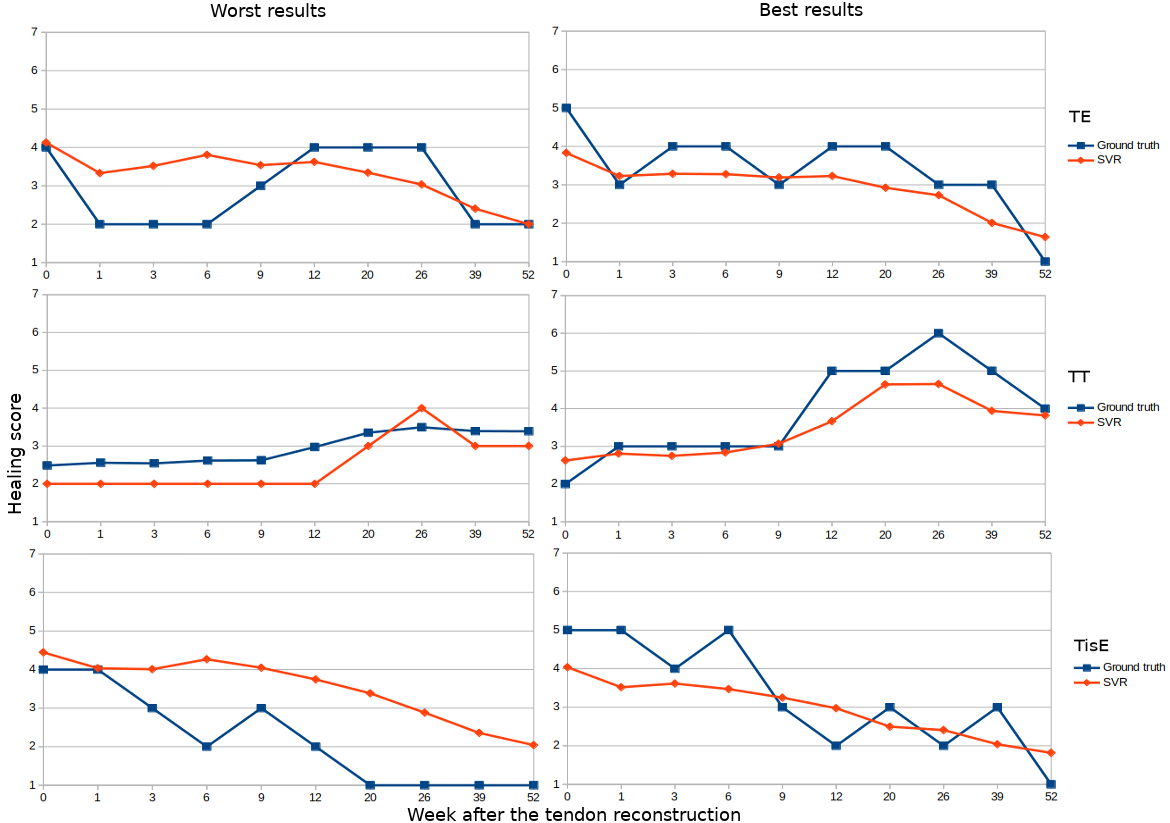}
\caption{Comparison of healing scores for TT, TE and TisE parameters.} \label{fig:healing-scores}
\end{figure*}

In Fig. \ref{fig:healing-scores} we show an example of the worst and best assessments in terms of the correlation for both intra- and extra-tendon parameters. The results are presented for the SVR, which resulted best in 5 scores in terms of the correlation while remaining competitive in other metrics.

We observe that in all cases the outcome of the rehabilitation and the starting point is assessed in a similar manner (with an absolute error below $1$ or approx. $1$ in one case). The level of 1 is also similar to the radiologist score uncertainty and should have minimal impact on clinical decisions. The assessment fluctuations during the healing process will be further discussed. 

\section{Discussion}

We show that our late fusion approach and the selection of appropriate deep learning and hand-crafted features improve the automatic assessment of the Achilles tendon healing. The process is affected by many disturbances like patient activity, diet and their obedience to the treatment prescription. According to the feedback provided by radiologists and medical professionals, extra-tendon conditions, especially edema, are more prone to these factors, hence explicitly incorporating the hand-crafted features that put emphasis on intra-tendon area results in an overall improvement.

The key to the boost in performance is to select only a small number of significant predictors from a mixture of hand-crafted and deep learning predictors. Haralick's features depend on pixel co-occurrence and are indicators of specific textural patterns. It is advantageous particularly in the case of the SCT, TE and TU. ROI statistics are particularly significant in terms of TT assessment, which resulted in the overall best scores for all of the computed metrics. On top of that, the inclusion of deep learning features after the principal component transformation allowed for the successful assessment of an extra-tendon parameter i.e. TisE. The other one, namely STE, resulted in a relatively low correlation, although still improved. This score doesn't rely on Haralick's and only indirectly incorporates ROI information, like area. The improvement mainly comes from the use of five DL features and not one like in the case of the baseline method.  
\vspace{-5px}
\section{Conclusions}

In this paper, we proposed a meta-regression model based on the late fusion of deep learning and hand-crafted features that improved the existing state-of-the-art automatic assessment of the healing Achilles tendon visible in the MRI studies. 

Furthermore, in-line with the achieved improvement we decreased the number of MRI protocols required for the approach from two to one, namely T2$^\ast$ GRE TE\_MIN. This directly translates to savings in time and costs, while the acquisition of data for our method takes approx. 5 minutes and for the previous approach approx. 12.

As for future work, we plan to automate the segmentation of the tendon region of interest (ROI), which currently remain manual. We have performed initial tests that indicate the possibility of employing fully convolutional networks for this task.

\subsection*{Acknowledgments}

{
The following work was part of {\it Novel Scaffold-based Tissue Engineering Approaches to Healing and Regeneration of Tendons and Ligaments (START)} project, co-funded by The National Centre for Research and Development (Poland) within STRATEGMED programme
(STRATEGMED1/233224/10/NCBR/2014).
}

%
%
%
\bibliographystyle{splncs04}
\bibliography{biblio}

\end{document}